\documentclass[aps,nofootinbib,superscriptaddress, showpacs,preprintnumbers,  nofootinbib,referee]{revtex4-1}
\usepackage{eurosym}
\usepackage{amsmath}
\usepackage{bm}
\usepackage{amsfonts}
\usepackage{amssymb}
\usepackage{graphicx}
\usepackage{mathtools}
\usepackage{physics}

\def\be{\begin{equation}}
\def\ee{\end{equation}}
\def\bea{\begin{eqnarray}}
\def\eea{\end{eqnarray}}

\newcommand{\f}[2]{\frac{#1}{#2}}

\begin{document}

\title{Matter really does matter, or Why $f(R,{\rm Matter})$ type theories are significant for gravitational physics and cosmology}

\author{Tiberiu Harko}
\email{tiberiu.harko@aira.astro.ro}
\affiliation{Department of Physics, Babes-Bolyai University, Kogalniceanu Street,
	Cluj-Napoca, 400084, Romania,}
\affiliation{Astronomical Observatory, 19 Ciresilor Street,
	Cluj-Napoca 400487, Romania}

\author{Miguel A. S. Pinto}
\email{mapinto@fc.ul.pt}
\affiliation{Instituto de Astrofísica e Ciencias do Espaco, Faculdade de Ciencias da Universidade de Lisboa, Edifício C8, Campo Grande, P-1749-016
Lisbon, Portugal}
\affiliation{Departamento de F\'{i}sica, Faculdade de Ci\^{e}ncias da Universidade de Lisboa, Edifício C8, Campo Grande, P-1749-016 Lisbon, Portugal}

\author{Shahab Shahidi}
\email{s.shahidi@du.ac.ir}
\affiliation{School of Physics, Damghan University, Damghan 36716-45667, Iran}

\begin{abstract}
In a recent paper (Lacombe, Mukohyama, and Seitz, JCAP {\bf 2024}, 05, 064 (2024)), the authors provided an in-depth analysis of a class of modified gravity theories, generally called $f(R,{\rm Matter})$ theories, which assume the existence of a non-minimal coupling between geometry and matter. It was argued that if the matter sector consists of Standard Model particles, then these theories suffer from the presence of ghosts, or are just scalar/vector-tensor theories. Hence, the relevance of these theories for cosmology was questioned. It is the goal of the present work to carefully analyze, discuss, and assess the line of arguments proposed in Lacombe et al. JCAP {\bf 2024}, 05, 064 (2024). After a qualitative critical discussion of the five general arguments proposed for the validity of a gravitational theory, we present the theoretical foundations of $f(R,{\rm Matter})$ theories, including their possible relations with quantum gravity, and discuss in detail the role of matter. The matter source discussed in Lacombe et al., consisting predominantly of a massless scalar field, is extremely restrictive, and rather irrelevant to cosmology and the description of the observational data. We also devote a detailed discussion of the problem of the energy scales of the $f(R,{\rm Matter})$ theories. To test the observational relevance of this type of theories we present the comparison of a simple theoretical model with a small set of observational data and with the $\Lambda$CDM paradigm.  We conclude by pointing out that the analysis of Lacombe et al., JCAP {\bf 2024}, 05, 064 (2024), even very useful for the understanding of some limited aspects of the $f(R,{\rm Matter})$ theories, and of their theoretical foundations, cannot be considered as a valid or definite criticism of these approaches to gravity.
\end{abstract}

\pacs{03.75.Kk, 11.27.+d, 98.80.Cq, 04.20.-q, 04.25.D-, 95.35.+d}
\date{\today }
\maketitle
\tableofcontents


\section{Introduction}

In an interesting and thought-provoking paper \cite{Muk}, the authors performed an extensive analysis of a particular class of modified gravitational theories, generically referred to as non-minimal geometry-matter coupling theories, with a special emphasis on two of them, $f\left(R, L_\text{m}\right)$ \cite{1} and $f(R, T)$ \cite{2} gravity theories. Following \cite{Muk}, in the subsequent analysis, we will refer to these theories, as well to their extensions, as $f(R,{\rm Matter})$ type theories. Some aspects of this class of theories have been presented rather extensively in the book \cite{3a}. For the sake of completeness, one could also mention some other gravitational theories involving non-minimal geometry-matter couplings, such as theories that couple matter and torsion \cite{tors}, and nonmetricity to the trace of the matter energy-momentum tensor \cite{nmetr1,nmetr2}.

Five relevant criteria for the validity of a generalization of Einstein's General Relativity (GR) were proposed, briefly discussed, and summarized in \cite{Muk}. These are stated as follows: 1. The matter sector containing the Standard Model (of elementary particles). 2. Not simply a modification of the matter sector. 3. Relevant energy scales and parameters. 4. Absence of ghosts. 5. Compatibility with observations. To these criteria, we would like to add one more, namely, 6. The existence of a possible relation with quantum gravity.

Based on the first five criteria, the authors of \cite{Muk} claim that the $f(R,{\rm Matter})$ theories are a) equivalent to $f(R)$ plus matter Lagrangian b) they only modify the matter sector with respect to GR or $f(R)$ c) they are just a rewriting of various other theories, and d) they are affected by the presence of ghosts and other instabilities within their domain of application to cosmology.

The authors of \cite{Muk} end their work by stating "As it is impossible to address every model within the $f(R, \rm{Matter})$ class, we encourage advocates of these models to examine whether specific instances somehow circumvent the problems discussed here. This paper aims to clarify why one should not expect such exceptions."

This is a very good suggestion, indeed, which is worth considering in detail. Nonetheless, before moving to the investigation of the many possible types of $f(R,{\rm Matter})$ theories following the above recommendation, and show that they are all inconsistent physically and theoretically (as claimed in \cite{Muk}), it would be important to clarify the meaning and the relevance of the criteria proposed and discussed by the authors of \cite{Muk} for these theories. Some of the points raised in \cite{Muk} were already considered in \cite{Ayuso} (Ref. [26] of \cite{Muk}), where the discussion mostly focused on the problem of the instabilities and the introduction of the trace of the matter energy-momentum tensor in the general formalism.

It is the goal of the present work to provide a critical analysis of the comments and results of \cite{Muk} and to give some relevant answers to the points raised in the work. In particular, we will point out some shortcomings in the physical and cosmological reasoning of \cite{Muk}, and we will try to shed some light on the correct interpretation of the geometry-matter coupling class of theories.

Moreover, we would like to firmly point out that, at this moment, we do not consider $f(R,{\rm Matter})$ type theories as the ultimate answer to all the problems of modern cosmology or to represent the final theory of gravity. Still, they may portray a (perhaps small) step toward the understanding of the structure and dynamical behavior of the gravitational field(s), its theoretical foundations, and its complex interaction with the cosmic environment. In addition, they may represent some attractive models for the interpretation and explanation of the present-day cosmological data.

The present paper is organized as follows. First of all, in Section~\ref{sect1}, we will critically assess, on a qualitative level, the validity of the five criteria introduced in \cite{Muk} as fundamental requirements for the $f(R,{\rm Matter})$ type theories. In particular, we will point out the limitations introduced concerning the definition of matter and the misunderstanding of the problem of the energy (correctly: length) scales. We will then proceed, in Section~\ref{sect2}, to the presentation of the theoretical foundations of $f(R,{\rm Matter})$ type theories. More specifically, we will point out the possible connection between this class of theories and quantum theories of gravity, and show how at least one simple theoretical model emerges from quantum considerations. We will also discuss other theoretical justifications of the $f(R,{\rm Matter})$ theories, and we will analyze on a more quantitative level the problem of matter, as it appears in the non-minimal geometry-matter coupling approaches to gravity, as well as  the important issue of ghost instabilities. The problem of the energy/length scales is discussed in Section~\ref{sect3}, where various aspects considered in \cite{Muk} are critically assessed. The observational evidence for the $f(R,{\rm Matter})$ theories is briefly addressed in Section~\ref{sect4} by considering a specific cosmological model, whose predictions are compared with a small set of observational data for the Hubble function, and with the predictions of the standard $\Lambda$CDM model. Finally, we conclude our paper with a general discussion and assessment of the paper \cite{Muk} in Section~\ref{sect5}.

\section{General arguments - a critical analysis}\label{sect1}

We will begin our discussion of the main points raised in \cite{Muk} by considering the (logical and physical) consistency  of the general arguments proposed in the analysis of the $f(R,{\rm Matter})$ type theories. The following arguments for the validity of gravitational and cosmological theories have been advanced, and discussed in some detail:
\paragraph{Criterion 1 - {\bf The matter sector containing the Standard Model.}} The authors of \cite{Muk} pointed out that the quantities related to the particle content of the theory, such as the matter Lagrangian, or the matter energy-momentum tensor, must belong to the Standard Model, thus being represented by photons, Higgs, etc. Moreover, in Eq. (2.9) of \cite{Muk}, an example is provided, illustrating how the matter Lagrangian should be constructed, and what it should contain, as follows
\begin{equation}\label{2.9}
\mathcal{L}_m=-\frac{1}{2}\nabla _{\mu}\phi\nabla^\mu \phi-m^2\phi^2+\lambda \phi^4 -F_{\mu \nu}F^{\mu \nu}+....
\end{equation}

Most of the discussions of various models and theories in \cite{Muk} are based on the simple assumption laid down in Eq. \eqref{2.9}.  However, the author's interpretation of the Standard Model in the astrophysical and cosmological context is at least strange since they seem to forget that the elementary particles of the Standard Model that are relevant for astrophysics and cosmology (in addition to photons and neutrinos, which are the two usual radiation components) are electrons and quarks, with the latter ones composing protons and neutrons. In fact, the matter we detect observationally in astrophysical and cosmological surveys consists of these types of basic particles (protons, neutrons, electrons), which we usually refer to as ordinary matter.

 However, as we will discuss in following, this ordinary matter should be described by an energy density and a thermodynamic pressure. This is simply because cosmology considers the universe as a whole, which responds naturally to the thermodynamics nature of the baryonic matter. {\it The main goal of the} $f(R,{\rm Matter})$ {\it type theories is exactly to point out the role of ordinary matter (protons, neutrons, hydrogen, helium, and so forth) in the cosmic dynamics}. Based on Criterion 1, astrophysics would be impossible since the structure of stars is mostly, and extremely successfully, investigated by considering protons, neutrons, electrons, etc., and not scalar or Higgs fields. Indeed, assuming that the Sun is constituted by protons and photons gives a pretty good description of its physics. Interestingly enough, the case of ordinary matter, as defined above, is not discussed at all in \cite{Muk}.
\paragraph{Criterion 2 - {\bf Not simply a modification of the matter sector}.} As such, this criterion is not physical, or cosmological, but it may be called methodological or perhaps taxonomical (taxonomy being the science of organizing things into groups or types). The authors of \cite{Muk} try to eliminate from the general class of $f(R,{\rm Matter})$ theories models in which only the matter part is modified, that is, theories of the type $f(R,{\rm Matter})=g(R)+h({\rm Matter})$, even that they clearly belong to the general class of $f(R,{\rm Matter})$ theories. At this point, the authors mention again the Standard Model of particle physics, consisting, in their view, only of photons, Higgs, etc., and claim that such a modification would contradict particle experiments, forgetting again that (ordinary) matter in cosmology consists of protons, neutrons, electrons, and so on. The statement that ``these theories cannot be seen as  a modification of the gravity sector" is incorrect. A correct statement would be: these theories cannot be seen as a {\it direct} modifications of the gravity sector, since {\it any modification of the matter source automatically influences gravity}. As a simple example, the standard general relativistic behavior of a polytropic gas with $p\propto \rho ^n$ is certainly different from the behavior of a gas satisfying a linear barotropic equation of state $p=(\gamma -1)\rho$. Hence, modifying the matter sector instantly modifies gravity.
\paragraph{Criterion 3 - {\bf Relevant energy scales and parameters}.} As a third criterion for a consistent gravitational theory, the authors require the theory to be consistent until a certain ultraviolet (UV) energy scale, which would fix the validity of the theory. The concept of UV energy scale, or generally an energy scale, borrowed from quantum field theory and high-energy particle physics, is not appropriate, or valid, in this context. In the classical gravitational theories, we do not consider energy scales, but length scales. From this perspective, we consider the Solar System scale, where standard GR is certainly valid, the galactic scale, where present-day science faces the dark matter issue, and the cosmological scale, where we are confronted with the dark energy and accelerating Universe  problems. Hence, the typical length scales for the $f(R,{\rm Matter})$ theories are the Solar System ($1.2\times 10^{-5}$ parsecs, the galactic (10-100 kpc), extragalactic (few Mpc to 100 Mpc), and the cosmological (3000-6000 Mpc) length scales. How to relate these length scales to some specific energy scales seems to be still an open problem.

Modified gravity theories have been specifically designed to describe physics on these length scales. Set of coupling parameters $\left(\lambda _1,\lambda _2,...\right)$ are indeed introduced in the models, but only in order to reduce them to GR in the limit $\lambda _i\rightarrow 0$. Furthermore, these coupling constants can generally be determined from observations. Another length scale on which $f(R,{\rm Matter})$ theories have been intensively investigated is the length scale of astrophysical compact objects (consisting of protons, neutrons, electrons, and so on), where the high matter density and the strong gravitational field may allow testing the effects of geometry-matter coupling. From this point of view, an important criterion for the validity of a $f(R,{\rm Matter})$ type theories would be that the same set of parameters describe the physics from the astrophysical scales to the cosmological ones, assuring, in this way, the uniqueness of the theoretical model.

\paragraph{Criterion 4 - {\bf  Absence of ghosts}.} The stability of perturbations about a Minkowski background in modified gravity theories is a fundamental test of the validity of the theories. The ghost instabilities may lead to potentially negative expressions of kinetic terms, involving propagating degrees of freedom.  Laplacian instabilities appear when the speeds of the propagation of the perturbations are not positive. The presence of instabilities can thus lead, for example,  to exponential growth rates. Thus the study of perturbations is fundamental for obtaining viable and self-consistent modified gravity theories, as well as realistic cosmological models.

The problem of the stability of modified gravity theories is also intimately related to the problem of matter. 
There are two main approaches to consider the matter fields in cosmology.
The first one is to borrow some fields from  particle physics with definite Lagrangian, as proposed in \cite{Muk}. In this case, the theory is reduced to a scalar/vector-tensor theory of gravity. These modifications of gravity theories in their matter sector have been intensively investigated in the literature, and can be used, for instance, to describe the early Universe, and the inflationary epoch. However, these theories are effectively scalar-tensor type theories and the stability issues should be examined in the context of field theories. This is not the approach for considering specially late time cosmology.

However, a different approach is used when  conceptualizing modified gravity theories with a non-minimal geometry-matter coupling. In this class of theories, it is generally assumed that the matter fields can be described, thermodynamically, by a perfect fluid, with a Lagrangian density of the form $L_m=-\rho$ or $L_m=p$. Specifically, as we have discussed before, {\it in} $f\left(R,{\rm Matter}\right)$ {\it theories the matter field are described thermodynamically}.

In thermodynamics, perfect fluids are locally described by several thermodynamic variables, $n$, $\rho$, $p$, $T$ and $s$, the particle number density, energy density, pressure, temperature and entropy per particle, respectively.
These variables are spacetime scalars, whose values represent measurements made in a rest frame of the fluid \cite{P2}.
The motion of a perfect fluid is described by the 4-velocity $u^\mu$. In short, a perfect fluid can be characterized by 5 of the above thermodynamic variables, namely, $\rho$, $p$ and $u^\mu$.

As a result, when we consider the matter field as a thermodynamics quantity, we treat the energy-density and pressure as two independent fields, not constructed from some scalar/vector fields. This is, in fact, more meaningful and reasonable in cosmology, since the detailed matter content of the Universe does not affect the cosmological behavior, which is determined by the averaged thermodynamic quantities, or by the thermodynamic behavior of the matter variables. This is our main reasoning of the absence of ghosts in $f\left(R,{\rm Matter}\right)$ theories. The ghost in \cite{1} appears because $\rho$ and $p$ are assumed to be constructed from the Kinetic/Potential terms of some scalar/vector field, which causes the theory to have more derivatives per field in each term, signaling the presence of ghost.

For example, when we consider an action of the form
\begin{align}
	S=\int d^4x\sqrt{-g}(\kappa^2 R+\alpha RT+L_m),
\end{align}
where $T$ is the trace of the matter energy-momentum tensor, and $\alpha$ is a coupling constant, the term $\alpha RT$ does not have more derivatives per fields, since $T$ consists of $\rho$ and $p$ which do not have any derivatives of other fields. $\rho$ and $p$ are taken to be dynamical fields, related to each other by some equation of state. This could be compared with the term $\alpha\phi R$ in scalar-tensor theories where $\phi$ is some scalar field which does not contain any derivative of other fields. As a result, $f\left(R,{\rm Matter}\right)$  theories do no suffer from ghost instabilities. However, we should mention that theories with derivative matter couplings should be analyzed more carefully from the above point of view \cite{derivativematter}.

To summarize the above brief discussion, the theories with geometry-matter couplings do not have a ghost instability problem because of their matter couplings, in contrast to the discussion in \cite{Muk}.

\paragraph{Criterion 5 - {\bf Compatibility with observations.}} The compatibility with the experiments/observations is the main criterion for the acceptance of any physical theory. In this regard, the $f(R,{\rm Matter})$ theories underwent an extremely detailed analysis at astrophysical, galactic, and cosmological scales, with a large variety of results obtained - too many to be discussed in the context of the present paper. Some investigations fully confirmed the theoretical predictions of the considered models, while others did find some discrepancies. Nevertheless, overall one could firmly claim that $f(R,{\rm Matter})$ type theories are at least a useful tool for the understanding of the cosmological dynamics, and have opened new windows on the understanding of dark matter and dark energy as geometric effects. On the other hand, the statement of the authors of \cite{Muk} that "Last but not least, we require that the theory
under consideration does not lead to observable quantities in contradiction with experiments." is rather incorrect, since in the fields of gravitational theories we are not dealing with experiments, but with observations (which are distinct from experiments). Secondly, it is difficult to understand what kind of elementary particle experiments could test modified gravity theories. The authors refer to experiments that may improve our knowledge of the matter Lagrangian (probably understood in terms of Higgs, scalar fields, photons, etc.), but such an improvement may not be directly relevant to the astrophysical and cosmological applications of the modified gravity theories. However, experiments that would certainly contribute to a better understanding of these theories, and of their astrophysical implications, may be related to nuclear physics experiments that could allow a better comprehension of the equation of state of dense nuclear matter (consisting of neutrons and protons), of the nuclear interactions, or of the formation and properties of the quark-gluon plasma. The experimental and theoretical improvements in the understanding of the properties of dense nuclear matter could certainly lead to some significant improvements in the modelling of the geometry-matter coupling in realistic astrophysical and cosmological systems.

As one can see from the above discussion, the arguments presented in \cite{Muk}, and considered by the authors of \cite{Muk} as a severe criticism of the $f(R,{\rm Matter})$ theories, even if (partially)  valid at a very general conceptual level, have severe shortcomings when applied to the context they discuss. There is a misrepresentation of the authors of \cite{Muk} of the concept of matter, as it appears in the astrophysics and cosmology of the $f(R,{\rm Matter})$ type theories. Moreover, the modification of the matter sector certainly affects, though indirectly, the gravitational one. And, more importantly, from the observational point of view, the $f(R,{\rm Matter})$ gravity still continue to be convincing alternatives to GR in the interpretation of the cosmological data.

\section{Theoretical foundations of $f(R,{\rm Matter})$ theories}\label{sect2}

In the present Section, we will review some of the theoretical sources, and justifications, of the $f(R,{\rm Matter})$ type theories, whose essential property is the presence of a non-minimal coupling between matter and geometry. We will show first that such a coupling has its origin in various approaches to quantum gravity, and that these theories may represent a phenomenological description of some quantum processes. We will also discuss the origin of the $f(R,{\rm Matter})$ theories as corresponding to the maximal extension of the Einstein-Hilbert action principle. Finally, we will consider, from a more quantitative point of view, the problem of matter, and of instabilities, as they appear from the perspective of the modified gravity theories with geometry - matter coupling.

\subsection{From quantum gravity to geometry-matter coupling}

The authors of \cite{Muk} did not mention/discuss any possible relation between quantum gravity theories and theories with geometry-matter coupling. However, we have added such a relation as the sixth criterion for the viability of a modified gravity theory. Indeed, while the quantization of gravity remains as one fundamental problem in theoretical physics with no particular convincing solution in sight, perhaps some hints on its nature and structure could be obtained at the level of some classical theories, and theories with geometry-matter coupling may belong to this class.  Hence, in the following, we are going to display some intricate relations between quantum gravity and geometry-matter couplings.

\subsubsection{Matter-geometry coupling from quantum fluctuations}

 Generally, by fully taking into account the nature of quantum mechanics, the general program for the quantization of gravity can be formulated as follows. In the standard approach of a quantum description, quantum operators must be associated to all physical (or, in the case of gravity, geometrical) variables. Hence, once the requirement of a full {\it non-perturbative quantum description} of gravity is adopted, the quantized Einstein gravitational field equations must be formulated in an operator form, given by \cite{re8,re9,re10}
\begin{eqnarray}
\label{einstein_quantum}
\hat{R}_{\mu\nu}-\frac{1}{2}\hat{g}_{\mu\nu}\hat{R}=\frac{8\pi G}{c^4}\hat{T}_{\mu\nu}.
\end{eqnarray}

The experimentally/observationally relevant information from the quantum gravitational Einstein equations in their operator form comes from taking their average over all possible products of the metric operators $\hat {g}\left(x_1\right)...\hat{g}\left(x_n\right)$ \cite{re8,re9,re10}. Hence, to obtain experimental predictions on quantum gravity we must solve the following set of equations for the Green functions $\hat{G}_{\mu\nu}$
\begin{eqnarray*}
\langle\Psi|\hat{g}(x_1)\hat{G}_{\mu\nu}|\Psi\rangle&=&\langle\Psi|\hat{g}(x_1) \hat{T}_{\mu\nu}|\Psi\rangle,\\
\langle\Psi|\hat{g}(x_1)\hat{g}(x_2) \hat{G}_{\mu\nu}|\Psi\rangle&=&\langle\Psi|\hat{g}(x_1)\hat{g}(x_2) \hat{T}_{\mu\nu}|\Psi\rangle,\\
\dots&=&\dots,
\end{eqnarray*}
where $|\Psi\rangle$ is a quantum state that is not necessary to be taken as the ordinary vacuum state. Unfortunately, the equations for the gravitational Green functions cannot be solved analytically, and thus the only possibility we have is the use of some approximate methods \cite{re8}-\cite{re10}. One interesting possibility proposed in \cite{re8} is related to the representation of the metric operator $\hat{g}_{\mu\nu}$  as the sum of two components, the average metric $g_{\mu \nu}$, and a fluctuating component $\delta\hat{g}_{\mu\nu}$.  Hence, we assume that the quantum metric can be decomposed as
$\hat{g}_{\mu\nu}=g_{\mu\nu}+\delta\hat{g}_{\mu\nu}$. Then, the quantum Einstein-Hilbert Lagrangian, $L_{\hat{g}}=\frac{1}{2k^2}\sqrt{-\hat{g}} \hat{R}$, can be approximated as
\begin{eqnarray}
\label{lag0}
L_{\hat{g}}=L_{\hat{g}}(g+\delta\hat{g})\approx L_{g}(g)+\frac{\delta L_{g}}{\delta g^{\mu\nu}} \delta \hat{g}^{\mu\nu},
\end{eqnarray}
where  the higher-order fluctuations have been neglected. Since $\langle\delta \hat{g}^{\mu\nu}\rangle\neq 0$, for the expectation value of $\frac{\delta L_{g}}{\delta g^{\mu\nu}}\delta\hat{g}^{\mu\nu}$ we find the expression \cite{re8, re11}
\be
\label{average}
 \left\langle \frac{\delta L_{g}}{\delta g^{\mu\nu}}\delta \hat{g}^{\mu\nu}\right\rangle=\frac{\delta L_{g}}{\delta g^{\mu\nu}}\langle \delta \hat{g}^{\mu\nu}\rangle=\frac{1}{2k^2}\sqrt{-g}G_{\mu\nu}\langle \delta \hat{g}^{\mu\nu}\rangle,
\ee
where $G_{\mu \nu} = R_{\mu \nu} -(1/2)g_{\mu \nu} R$ denotes the Einstein tensor in classical geometry. Hence, with the use of Eq. \eqref{average}, for the average value of the Lagrangian (\ref{lag0}) we obtain
\begin{eqnarray}
\label{lag00}
\langle L_{\hat{g}}\rangle\approx \frac{1}{2k^2}\sqrt{-g} \big[R+G_{\mu\nu}\langle\delta \hat{g}^{\mu\nu}\rangle \big].
\end{eqnarray}

We can also approximate the quantized Lagrangian of the matter, $L^{\hat{g}}_{\rm m} = \sqrt{- \hat{g}} L_\text{m}(g)$,  as \cite{re8}
\begin{eqnarray}
\label{lagm}
L^{\hat{g}}_{\rm m}= L^{\hat{g}}_{\rm m}(g+\delta\hat{g})\approx \sqrt{-g}L_{\rm m}(g)+\frac{\delta (\sqrt{-g}L_{\rm m})}{\delta g^{\mu\nu}}\delta \hat{g}^{\mu\nu}.
\end{eqnarray}
Thus,  the expectation value of the quantum matter Lagrangian (\ref{lagm}) can be written as \cite{re8}
\begin{eqnarray}
\label{lagm0}
\langle L^{\hat{g}}_{\rm m} \rangle \approx \sqrt{-g}\big[L_{\rm m}-\frac{1}{2}T_{\mu\nu}\langle\delta \hat{g}^{\mu\nu}\rangle\big],
\end{eqnarray}
where  by $T_{\mu\nu}=-2\delta(\sqrt{-g}L_{\rm m})/(\sqrt{-g}\delta g^{\mu\nu})$ we have denoted the classical matter energy-momentum tensor.
Therefore, the averaged total quantum modified Lagrangian, containing both the gravitational and the matter sectors,  can be written as (for the computational details see \cite{re8,re9,re10})
\begin{eqnarray}
\label{lag}
L=\frac{1}{2k^2}\sqrt{-g} \big[R+G_{\mu\nu}\langle\delta \hat{g}^{\mu\nu}\rangle \big]+\sqrt{-g}\big[L_{\rm m}-\frac{1}{2}T_{\mu\nu}\langle\delta \hat{g}^{\mu\nu}\rangle\big].\nonumber\\
\end{eqnarray}

It is easy to see that metric variation $\langle\delta \hat{g}^{\mu\nu}\rangle$ has the same symmetry properties as the metric $g^{\mu\nu}$. Hence, in a first order of approximation, we assume for the quantum metric fluctuation tensor $\langle\delta \hat{g}^{\mu\nu}\rangle$ the simple form $\langle\delta \hat{g}^{\mu\nu}\rangle=\alpha g^{\mu\nu}$ \cite{re8} (but other choices are also possible).  Therefore, the gravitational Lagrangian (\ref{lag}) becomes \cite{re11}

\begin{eqnarray}\label{act}
L=\left[\frac{1}{2k^2}(1-\alpha)R+\left(L_{\rm m}-\frac{1}{2}\alpha T\right)\right]\sqrt{-g},
\end{eqnarray}
where $T=g^{\mu\nu}T_{\mu\nu}$ is the trace of the energy-momentum tensor. It is easy to verify that the Lagrangian \eqref{act} is nothing but a particular example of the class of the additive $f(R,T)$ modified gravity theories \cite{2}. Despite its simplicity, the Lagrangian \eqref{act} has a purely  quantum origin, and other forms, involving more complicated geometry-matter couplings can be constructed as well. Thus, specific cases within $f(R,T)$ gravity theory, or, more generally, $f(R,{\rm Matter})$ type theories could lead, even considered at a first-order phenomenological level, to the understanding of some sophisticated aspects of quantum gravity.

\subsubsection{Coupling matter and geometry at the quantum level}

Modified gravity theories with non-minimal geometry-matter coupling may also be formulated directly at the semi-classical level. Such an approach, having many affinities with the purely classical modified gravity theories with non-minimal coupling,  was proposed in \cite{Kibble}. In particular, this theoretical model assumes a non-minimal coupling between the quantum fields and the classical scalar curvature $R$. The action, explicitly including the geometry-quantum matter coupling term, is given by
\be\label{coupl}
\int{G\left(\left<g(\phi)\right>_{\Psi}\right) R \;\sqrt{-g}\;d^4x},
\ee
where $G$ and $g$ are arbitrary functions, and the average of $g$ is obtained as $\left<g(\phi)\right>_{\Psi}=\left<\Psi (t)\right|g[\phi (x)]\left|\Psi (t)\right>$. This geometry-matter coupling term leads to a new Hamiltonian $H(t)$ in the Schr\"odinger equation
\be\label{Sch}
i|\dot{\Psi}(t)\rangle=\hat{H}(t)\left|\Psi (t)\right>-\alpha (t)\left|\Psi (t)\right>,
\ee
where by $\alpha (t)$ we have denoted a Lagrange multiplier, given by \cite{Kibble}
\be
\hat{H}(t)\rightarrow \hat{H}_{\Psi}=\hat{H}(t)-\int{N G'\left(\left<g(\phi)\right>_{\Psi}\right)g(\phi)\sqrt{\Sigma}\; d^3\xi},
\ee
where $N$ denotes the lapse function, while  are the intrinsic space-time coordinates.  The $\xi ^i$ coordinates satisfy the condition that the normal vector to a space-like surface is time-like on the entire space-time manifold. The scalar function $\Sigma $ is defined as $\Sigma = {\rm det} \;\Sigma _{rs}$, where $\Sigma _{rs}$ is the metric induced on a space-like surface $\sigma( t )$,  which has the property of globally slicing the space-time manifold into
space-like surfaces. Then, by using the coupling (\ref{coupl}) between the quantum fields and the classical geometry, the effective semiclassical Einstein equations can be obtained as \cite{Kibble}
\bea\label{123}
R_{\mu \nu}-\frac{1}{2}Rg_{\mu \nu}=16\pi \tilde{G}\Big[\langle \hat{T}_{\mu \nu}\rangle _{\Psi}+G_{\mu \nu}G- \nabla _{\mu}\nabla _{\nu} G+
g_{\mu \nu}\Box G\Big],
\eea
where by $\tilde{G}$ we have denoted the classical gravitational constant.

From Eq.~(\ref{123}) it follows that the matter energy-momentum tensor is not conserved,  $\nabla _{\mu}\langle \hat{T}^{\mu \nu}\rangle _{\Psi}\neq 0$. Hence, Eq.~(\ref{123}) describes a particle creation process, in which the energy is transferred from geometry to the matter fields, and this process is fully determined by the quantum structure of the space-time. Indeed, Eq.~(\ref{123}) gives us an effective description of the quantum processes in a gravitational field, in the presence of a geometry-quantum matter coupling, and thus gives important insights into possible quantum energy transfer and particle creation processes.

To conclude our present discussion, in the effective quantum gravity models we have presented, the energy-momentum tensor of the matter is not conserved. The non-conservation of matter energy-momentum is one of the most important properties of theories with a non-minimal geometry-matter coupling. Thus, these types of theories allow for the possibility of an energy transfer between space-time and matter.  Therefore, it is plausible to propose that the physical origins of the modified gravity theories with a non-minimal geometry-matter coupling, including matter creation processes, may be traced back to the physical properties of the quantum fields evolving in a curved space-time geometry. In the following, we will consider another source of theories with geometry-matter coupling, the problem of matter, as well as other issues raised in \cite{Muk}.

\subsection{Theoretical curiosity, matter, and its role in gravity}

We will continue our analysis of the fundamentals of the $f(R,{\rm Matter})$ gravity by pointing out another source of this type of theories, which we would like to call theoretical curiosity.  Then, we will address the problem of matter, as it appears in \cite{Muk} and in the gravitational theories with a non-minimal geometry-matter coupling.

\subsubsection{Maximal extension of the Einstein-Hilbert Lagrangian}

First, let us consider the well-known Einstein-Hilbert Lagrangian, $L=R/2\kappa ^2+L_\text{m}$. It is an extremely powerful, elegant, and simple Lagrangian, possessing an  additive Abelian algebraic structure. The Einstein-Hilbert Lagrangian is built up from two physical quantities, forming the set (in the proper mathematical sense) $\{R,L_\text{m}\}$. At the level of theoretical curiosity (which is essential in science), we have the freedom to ask the following question: what is the maximal extension, in a mathematical sense, of the Einstein-Hilbert Lagrangian? In the following, we use the term ``maximal extension'' in a strict mathematical (and logical) sense as referring to sets. In this interpretation, we define the maximum of an {\it ordered set}  $U=\{a_1,a_2,...,a_n,...\}$ {\it as an element} $a_{n+1}\in U$ {\it that is followed by no other}.

For example, beginning with the set $\{R,L_\text{m}\}$, we can construct the maximal set ${\rm GTGRL_m}=\Big\{{\rm Grav.\; Theor.\; \left(R,\;L_\text{m}\right)}\Big\}$ (generalized theories of gravity of the $R,L_m$ type) as given by
\begin{align}
{\rm GTGRL_m}&=\Big\{{\rm Grav.\; Theor.\; \left(R,\;L_\text{m}\right)}\Big\}
\nonumber\\&=\Big\{R+L_\text{m}, f(R)+L_\text{m}, R+f\left(L_\text{m}\right), f(R)+f\left(L_\text{m}\right),f(R)f\left(L_m\right)...,f\left(R,L_\text{m}\right)\Big\}.
\end{align}
The set ${\rm GTGRL_m}$ obtained in this way is finite, i.e. it has a finite number of elements.
Moreover, in this set-theoretical framework, the $f\left(R,L_\text{m}\right)$ gravity theory represents the maximal extension of the set of all gravitational theories built up in a Riemann geometry, and with the gravitational Lagrangian depending on the Ricci scalar $R$ and the matter Lagrangian $L_m$ only.

In a formal sense, the construction presented above can be easily generalized by considering some other sets of physical relevance. In the $f(R,T)$ modified gravity theory \cite{2}, the gravitational Lagrangian is defined by an arbitrary function of the Ricci scalar $R$ and of the trace of the energy-momentum tensor $T$. This theory can also be interpreted as the maximal extension of the gravitational theories constructed from the elements of the finite set $\big \{R,T\big\}$, leading to the set of ${\rm GTGRT}=\big\{{\rm Grav.\; Theor.\; \left(R,\;T\right)}\big\}$ gravitational theories, with the elements of the set constructed in a similar way as for the $f\left(R,L_\text{m}\right)$ gravity theories. This approach can be extended to include all the possible extensions of the Einstein-Hilbert Lagrangian. In fact, many such generalizations have been considered, based mostly on physical and cosmological considerations.  In the $f\left(R,T,R_{\mu \nu}T^{\mu \nu}\right)$ gravity theory \cite{4a}, a non-minimal coupling of matter to geometry is introduced, with the gravitational Lagrangian given by an arbitrary function of the Ricci scalar $R$, of the trace of the matter energy-momentum tensor $T$, and of the contraction of the Ricci tensor with the matter energy-momentum tensor. In this case, the geometry-matter coupling is also extended at the level of the Ricci tensor and not limited to the Ricci scalar.

Therefore, from at least the point of view of theoretical curiosity, and the point of view of the search for the limits of the natural laws,  the question of the investigation of the astrophysical and cosmological consequences of the maximal extensions of the Einstein-Hilbert Lagrangian is a valid one.  And of course an independent analysis of all the elements of the sets ${\rm GTGRL_m}$ or ${\rm GTGRT}$ from a physical point of view is also possible, and perhaps even necessary.

However, even from a purely mathematical perspective, the theoretical set extension of the Hilbert-Einstein Lagrangian may be justified, from a physical point of view the situation may be quite different. There are at least two reasons for this. First, the various elements of the above-considered sets may not be independent, and, thus, the element $R+L_\text{m}$ may be identical, or equivalent in some sense, to the term $R+f\left(L_\text{m}\right)$, let us say. This problem appears mainly at the level of matter, and this is an important point raised in \cite{Muk}. Secondly, it is important to find some physical (theoretical, observational, or experimental) criteria, which could reduce the number of the elements of the sets of the extended gravitational theories - in the ideal case to only one element.

In the next Subsection, we will analyze in detail the problem of the matter, and we will clarify the physical usage of the term matter in modified gravity theories, as compared to the usage of this term in \cite{Muk}.

 \subsection{Modified gravity and matter}

 The definition of matter in \cite{Muk} is rather restrictive, and considers matter mostly in the form of Standard Model particles, concentrating essentially on the presence of scalar fields, Higgs bosons, photons, etc. However, some of these forms of matter are either not applicable or irrelevant to the cosmology of geometry-matter coupling theories, as we are going to see now.

 \paragraph{Scalar fields in $f(R,\text{Matter})$ theories.} It is true that if one assumes matter to be represented by a scalar field only, $f\left(R,L_\text{m}\right)$ or $f(R,T)$ theories automatically reduce to a scalar-tensor theory. But this case is of no interest in the class of theories we are considering.
 In $f(R,T)$ gravity, or any other $f(R,\text{Matter})$ theory, one typically considers a scalar field in the matter sector for two distinct investigations: to seek for inflationary scenarios in the very early universe, or to explore self-gravitating systems, such as, for example, boson stars.

On the other hand, {\it the cosmological role of scalar fields in} $f(R,\text{Matter})$ {\it theories is minimal, if any}. In the non-minimal geometry-matter coupling theories, one of the main goals is to explain the late-time cosmic acceleration without any other component than the usual ordinary baryonic matter (composed of protons, neutrons, electrons, etc.) considered to be a perfect fluid, from a thermodynamic point of view. This is contrary to what happens in other cosmological theories, like, for example, in the quintessence models, in which it is the scalar field(s) that provide such an accelerated expansion, and are the source of dark energy. Indeed, {\it one important theoretical/aesthetic advantage of the non-minimal coupling theories is the evident reconsideration of the role of the baryonic content of the universe in the cosmological evolution, which is almost totally neglected in the dark energy-type theories}.

Furthermore, it is important to note that {\it free scalar fields as such have not been detected yet in the universe, at a cosmological or an astrophysical level}. The Higgs boson is also not directly detectable in the astrophysical or cosmological settings. Hence, as we have already mentioned,  {\it in the modified gravity theories with geometry-matter coupling, the concept of matter is constructed around particles like protons, neutrons, electrons, positrons, etc., which carry what we usually call the mass of the stars, galaxies, and of the entire universe}. The relevance of the particle accelerator experiments for the understanding of the matter couplings, as mentioned in \cite{Muk}, is thus rather doubtful, since all the interactions, including the gravitational ones, are carried out by the above-mentioned particles, whose physical properties we assume to be (relatively) well-known. Hence, $f(R,\text{Matter})$ {\it type theories generally do not need scalar fields in their cosmological approach}. The matter in $f(R,\text{Matter})$ theories is not some field theoretical dynamical scalar/vector field. It represents baryonic matter derived from the matter Lagrangian $L_m$ which is treated thermodynamically in the standard GR and hence modified gravities.

\paragraph{$f(R,\text{Matter})$ theory and scalar field models - further aspects.}
There is also another point that makes the difference between $f(R,\text{Matter})$ theory and the scalar field based approach more clear. As is well-known in particle physics, we can treat scalar/vector fields as thermodynamic quantities with definite energy-density and pressure. For instance, in the case of a canonical scalar field
\begin{align}
	S=\int d^4x\sqrt{-g}\left(-\frac12\partial_\mu\phi\partial^\mu\phi+V(\phi)\right),
\end{align}
one can define the energy-density, pressure and 4-velocity as
\begin{align}
	\rho&=-\frac12\partial_\alpha\phi\partial^\alpha\phi-V,\nonumber\\
	p&=-\frac12\partial_\alpha\phi\partial^\alpha\phi+V,
\end{align}
and
\begin{equation}
	u_\mu=\frac{\partial_\mu\phi}{\sqrt{-\partial_\alpha\phi\partial^\alpha\phi}},
\end{equation}
respectively. With the above definitions, the energy-momentum tensor of a scalar field
\begin{equation}
	T_{\mu\nu}=-\frac12\partial_\alpha\phi\partial^\alpha\phi g_{\mu\nu}+Vg_{\mu\nu}+\partial_\mu\phi\partial_\nu\phi,
\end{equation}
can be written as a perfect fluid energy-momentum tensor
\begin{align}
	T_{\mu\nu}=(\rho+p)u_\mu u_\nu+pg_{\mu\nu}.
\end{align}

This procedure can not be done inversely. In other words, we can not describe all the properties of a perfect fluid with one or more scalar/vector fields (we have argued that a perfect fluid need two scalars plus a 4-velocity field). This can also be seen from the fact that for the scalar field, only $L_m=p$ gives a correct result and $L_m=-\rho$ gives a wrong answer. In contrast, for a thermodynamic perfect fluid, both $L_m=p$ and $L_m=-\rho$ gives the same and also correct results \cite{secondvariation}.

\subsubsection{The definition of matter.}  We would like to reiterate that in $f(R,\text{Matter})$ type theories, by matter we usually understand a system of interacting particles (protons, neutrons, electrons, etc.), all belonging fundamentally to the Standard Model of particle physics. The nature, structure, and dynamics of the matter systems are determined by the fundamental macroscopic (or microscopic) laws of nature. Moreover, in the class of modified gravity theories we are considering, we usually assume matter to be in a fluid form. From the known laws of physics, one can obtain the physical and thermodynamic parameters of the matter fluids, such as energy density,  pressure, temperature, specific enthalpy, and so forth. However, in this context, it is also important to define how the physical quantities considered in these theories are obtained. In a very general sense, the basic physical (thermodynamic)  quantities that are considered in the $f(R,{\rm matter})$ type theories are the quantities that can be obtained from the known microscopic distribution functions of the particles (Boltzmann, Fermi-Dirac, Bose-Einstein,  etc.) \cite{Lan}. The properties of a fluid system can also be described, at least in some order of approximation, by a variational approach, which also allows the calculation of the matter energy-momentum tensor. However, the variational approach to fluid dynamics faces a very serious challenge, related to the degeneracy in the choice of the matter Lagrangian: $L_\text{m}=-\rho$ or $L_\text{m}=p$ lead to the same equations of motion, and energy-momentum tensor, a situation which has fundamental implications on the $f(R,{\rm Matter})$ type theories.

\paragraph{Gravitational effects on the thermodynamic quantities.} The important role of the interaction between matter and gravity already appears at the microscopic level when calculating the thermodynamic properties of matter.  The gravitational field modifies the distribution function of the particles at the microscopic level, as well as the total energy of the system, a situation that becomes extremely important for matter in strong gravitational fields.  In the standard approach for obtaining the equation of state of matter, when gravitational effects are included in the calculations, usually standard GR is used. Hence, it follows that the effects of GR, and through it of the gravitational interaction, may appear (somehow hidden) in the equation of state of high-density matter.

This fact has important implications on modified gravity theories since using a standard matter equation of state based on GR could lead to serious errors \cite{106}. For the applications of modified gravity theories, be they astrophysical or cosmological, it is important to use equations of state derived from particle and nuclear physics, which are independent of GR. A family of equations of state, independent of the gravitational theory,  was obtained in \cite{106} from a first principles approach, including causality, thermodynamic stability, and perturbation theory. These equations of state can be thus successfully  used either to constrain modified gravity theories or for the astrophysical or cosmological applications of these theories. However, the equations of state used in most of the cosmological and astrophysical investigations are obtained in the Minkowski space-time.

Indeed, it is important to understand that modified gravity has an important effect on the matter equation of state, as has been explicitly shown in several studies devoted to this problem. In the modified gravity theory with the Lagrangian $L_g = (R - 2\Lambda) /16\pi G + \alpha R^2 + \beta R_{\mu \nu}R^{\mu \nu} + \gamma R_{\mu \nu \tau \sigma}R^{\mu \nu \tau \sigma}$, where $\alpha$, $\beta $, $\gamma$ are constants, the thermodynamic properties of an ideal quantum gas were investigated in \cite{107}. An extended polytropic equation of state  $p = K\rho^\gamma +\epsilon \rho ^{3/2}$, where $K$, $\gamma$, and $\epsilon$ are constants, was found in the Eddington-inspired Born Infeld modified gravity \cite{108}. The modification of the equation of state is due to the gravitational backreaction on the particles.  The effects of the gravitational time dilation must also be included in the equations of state which are obtained in a curved geometry \cite{110,111}. For a Fermi gas, the equations of state were derived in the Palatini $f(R)$ gravity by considering both the relativistic and the nonrelativistic limits in Ref. \cite{112}.

The main conclusion of this brief discussion on the matter equation of state is the following: in order to obtain a consistent description of the astrophysical and cosmological properties of the physical systems, one must take into account the considered theory of gravity also at a microscopic level, and, therefore, a first principle approach is required for obtaining the properties of the high-density matter.

\subsubsection{The matter Lagrangian in modified gravity theories.} Similar results as discussed above on the effects of the modified gravity theories on the matter content do also appear when considering the matter Lagrangian. As shown in \cite{4}, the  matter energy-momentum in the $f(R,{\rm Matter})$ modified gravity theories has a different structure than the general-relativistic energy-momentum tensor, and it contains supplementary terms, which depend on the matter equation of state. To illustrate this result let us consider the $f(R,{\rm Matter})$ type modified gravity theory with the following action \cite{Bertolami:2007gv}
\begin{equation}
S=\int \left(\frac{1}{2}f_{1}(R)+\left[ 1+\lambda f_{2}(R)\right]
 L_{m}\right) \sqrt{-g}\;d^{4}x~,
\end{equation}
where $f_{i}(R)$ (with $i=1,2$) are two arbitrary functions of the Ricci scalar $R$, $L_{m}$ is the matter Lagrangian, and $\lambda$ is a coupling parameter. When $\lambda=0$, we recover the $f(R)$ gravity theory.  If the fluid is barotropic, one can take  $L_{m}=L_{m}\left( \rho \right) $. Then, the matter energy-momentum tensor becomes
\begin{equation}
T^{\mu \nu }=-\rho \frac{dL_{m}}{d\rho }u^{\mu }u^{\nu }+\left( L_{m}-\rho
\frac{dL_{m}}{d\rho }\right) g^{\mu \nu }, \label{tens}
\end{equation}
where $u^\mu$ is the fluid four-velocity. By using the equation of motion of massive particles
\begin{equation}
\frac{d^{2}x^{\mu }}{ds^{2}}+\Gamma _{\nu \lambda }^{\mu }u^{\nu }u^{\lambda
}+\left( u^{\mu }u^{\nu }-g^{\mu \nu }\right) \nabla _{\nu }\ln \sqrt{Q}=0,
\end{equation}
where we have denoted $\sqrt{Q}=\left[ 1+\lambda f_{2}(R)\right] \left(dL_{m}\left( \rho \right)/d\rho \right)$,
one can obtain that the matter Lagrangian for this modified gravity theory has the form \cite{4}
\begin{equation}\label{Lm}
L_{m}\left( \rho \right) =-\rho \left[ 1+\Pi \left( \rho \right) \right]
+\int_{p_{0}}^{p}dp,
\end{equation}
where $\Pi \left( \rho \right) =\int_{p_{0}}^{p}{\left(dp/d\rho\right)d\rho}$. The corresponding energy-momentum tensor is given by
\begin{equation}
T^{\mu \nu }=\left\{ \rho \left[ 1+\Phi \left( \rho \right) \right] +p\left(
\rho \right) \right\} u^{\mu }u^{\nu }+p\left( \rho \right) g^{\mu \nu },
\label{tens1}
\end{equation}
where we have denoted $\Phi \left( \rho \right) =\int_{\rho _{0}}^{\rho }\frac{p}{\rho ^{2}}d\rho
=\Pi \left( \rho \right) -\frac{p\left( \rho \right) }{\rho }$.

Therefore, it follows that the matter Lagrangian is both gravitational theory and model dependent.  By assuming a linear barotropic equation of state  $p=\left( \gamma -1\right) \rho $, $\gamma =$ constant, $%
1\leq \gamma \leq 2$, for the matter Lagrangian we find the expression
\begin{equation}
L_{m}\left( \rho \right) =-\rho \left( 1+\left(
\gamma -1\right) \left[  \ln \left( \frac{\rho }{\rho _{0}}\right) -1 %
\right] \right),
\end{equation}
 and $\Phi \left( \rho \right) =\left( \gamma -1\right)
\ln \left( \rho /\rho _{0}\right) $, respectively. For a polytropic equation of state, $p=K\rho ^{1+1/n}$, with $K,n=$constant, we obtain
\begin{equation}
L_{m}\left( \rho \right) =-\rho -K\left(\frac{n^{2}}{n+1}-1\right)
\rho ^{1+1/n},
\end{equation}
and $\Phi \left( \rho \right) =Kn\rho ^{1+1/n}=np\left( \rho
\right) $.  Finally, for the case of the ideal gas equation of state $%
p=k_{B}\rho T/\mu $, where $k_{B}$ is Boltzmann's constant, $T$ is the
temperature, and $\mu $ is the mean molecular weight, the matter Lagrangian is given by
\begin{equation}
L_{m}\left( \rho \right) =-\rho \left( 1+\frac{k_{B}T}{\mu }\left[
\ln \left( \frac{\rho }{\rho _{0}}\right) -1\right] \right) +p_{0}.
\end{equation}

By taking into account that  $L_\text{m}^{(0)}=-\rho$ is the Lagrangian of the matter with the effect of gravity on the thermodynamic properties of matter neglected, all the above expressions of the matter Lagrangian can be written concisely $L_\text{m}=f(L_\text{m}^{(0)})$, or as $f\left(L_\text{m}\right)$, where  $L_\text{m}=-\rho$.

Hence, the theories of the type $R+f\left(L_\text{m}\right)$ or $f(R)+f\left(L_\text{m}\right)$ are completely justified physically, contrary to the claim in \cite{Muk}. Indeed, these theories do not modify the gravitational sector but modify the matter one, by taking explicitly into account the effect of gravity and curved space-time on the matter equation of state. Of course, a first principle approach for these types of models is not used in all investigations, and sometimes an effective or phenomenological approach may be used, by simply postulating an analytical form for $f\left(L_\text{m}\right)$. Nonetheless, one should clearly point out that the thermodynamic quantities (temperature, density, pressure, etc.) obtained by taking into account the effects of the gravitational field on the classical or quantum distribution functions of particles are the real physical quantities, and thus they are the basic physical thermodynamic variables that must be used for the description of the gravitational processes in the presence of matter.

\subsubsection{Instabilities in $f(R, {\rm Matter})$ theories-the Dolgov-Kawasaki instability}

Beside consistency with the astrophysical and cosmological observations, including the Solar System tests, any gravitational theory should be stable against classical and quantum fluctuations. An important instability of modified gravity theories is the Dolgov-Kawasaki instability \cite{DK, DK1,DK1a, DK1b,DK1c,DK1d}, which we shall briefly discuss in the following, by relying on the presentation in  \cite{4a}.

Let us assume that, in order to be consistent with the Solar System tests, the Lagrangian of the $(R, {\rm Matter})$ theory can be represented as
\be
f(R,T,R_{\mu\nu}T^{\mu\nu})=R+\epsilon\Phi\left(R,T,R_{\mu\nu}T^{\mu\nu}\right),
 \ee
where $\epsilon$ is a small parameter. Following the approach pioneered in \cite{DK}, we expand the space-time quantities around a constant curvature background with geometrical and physical parameters $\left(\eta _{\mu \nu},R_0, T_{\mu \nu }^0, T_0,L_0\right)$, thus obtaining
\be\label{E1}
R_{\mu\nu}=\f{1}{4}R_0\eta_{\mu\nu}+R^1_{\mu\nu}, \;\;  R=R_0+R_1,
\ee
\be\label{E2}
T_{\mu\nu}=T^0_{\mu\nu}+T^1_{\mu\nu},\;\;  T=T_0+T_1,\;\;
 L_m=L_0+L_1,
\ee
where the metric tensor is represented as
$
g_{\mu\nu}=\eta_{\mu\nu}+h_{\mu\nu}
$, with $\eta _{\mu \nu}$ denoting the local Minkowskian metric. In the expansions (\ref{E1}) and (\ref{E2}) we have introduced two types of approximations, as pointed out, for example, in \cite{DK1}. The first approximation is the adiabatic expansion around a constant curvature space. This approximation is justified for time scales much shorter than the Hubble time. The second approximation used in Eqs.~(\ref{E1}) and (\ref{E2}) is a local expansion in the very small regions of the space-time, which are assumed to be locally flat. These approximations have been considered in detail in the study of the instabilities of the $f(R)$ type gravity theories \cite{DK,DK1}, and further extended to $(R, {\rm Matter})$ theories in \cite{4a}.
The Lagrangian density  $f\left(R,T,R_{\mu\nu}T^{\mu\nu}\right)$ can be also expanded as
\bea
f(R,T,R_{\mu\nu}T^{\mu\nu})
&=&R_0+R_1+\epsilon\bigg[\Phi(0)+\Phi_R(0)R_1
   +\Phi_T(0)T_1+\Phi_{RT}(0)\left(\f{1}{4}R_0T^1+R^1_{\mu\nu}T_0^{\mu\nu}\right)\bigg]\nonumber\\
&=&R_0+\epsilon\Phi(0)+\big[1+\epsilon\Phi_R(0)\big]R_1+H^{(1)},
\eea
where $(0)$ denotes the computation of the function at the background level, and for simplicity we have defined the first order quantity $H^{(1)}$ as
 $
 H^{(1)}=\epsilon\left[\Phi_T(0)T_1+\Phi_{RT}(0)\left(\f{1}{4}R_0T^1+R^1_{\mu\nu}T_0^{\mu\nu}\right)\right].
 $
 We then obtain

In the limit considered, one may write $\Box=-\partial_t^2+\nabla^2$, thus obtaining
\be\label{}
T_0^{\alpha\beta}\nabla_\alpha\nabla_\beta R_1=T^{00}_0 \ddot{R}_1+T_0^{ij}\partial_i\partial_j R_1.
\ee
One can then rewrite the above equation as
\be
\ddot{R}_1+U_{eff}^{ij}\nabla_i\nabla_j R_1+m_{eff}^2 R_1 = H_{eff},
\ee
where we have defined the quantity $U_{eff}^{ij}$ as
\be\label{DT}
U_{eff}^{ij}=\f{\big(3\epsilon\Phi_{R,R}(0)+\f{1}{2}\epsilon T_0\Phi_{RT,R}(0)\big)\delta^{ij}+\epsilon T_0^{ij}\Phi_{RT,R}(0)}{T_0^{00}-3\epsilon\Phi_{R,R}(0)-\f{1}{2}\epsilon T_0\Phi_{RT,R}(0)}.
\ee
For the explicit definitions of the effective mass  $m_{eff}$ and of $H_{eff}$ see \cite{4a}. The dominant term in Eq.~(\ref{DT})  is $1/\big[3f_{RR}(0)+(\f{1}{2}T_0 -T_0^{00})f_{RT,R}(0)\big]$, and hence one can formulate the condition of the avoidance of the Dolgov-Kawasaki instability in the $f(R,T,R_{\mu\nu}T^{\mu\nu})$ theory as
\be
3f_{RR}(0)-\left(\rho_0-\f{1}{2}T_0 \right)f_{RT,R}(0)\geq0,
\ee
where by $\rho _0$ we have denoted the background energy density of the matter \cite{DK}. It is important to note  that the condition for the stability does not depend on the derivative of the function $f$ with respect to $T$. {\it Therefore, we obtain the important result that the DK stability condition for the case of the} $f(R,T)$ {\it modified gravity theory is the same as in} $f(R)$ {\it gravity}. However, the stability condition is modified in the case of $f(R,T,R_{\mu\nu}T^{\mu\nu})$ theory.

\section{Energy scales}\label{sect3}

Another interesting issue raised in \cite{Muk} is questioning the existence of an energy scale in which the effects of the non-minimal geometry-matter coupling become dominant. Such a question is a very important, and valid one. However, we have already clearly mentioned that in $f(R,{\rm Matter})$ type theories the fundamental concept is the length scale, and not the energy scale. The relations between these two scales are at least elusive. But, since the authors of \cite{Muk} have devoted a lot of discussions of this issue, we will also briefly consider the main points of their analysis. Still, we are convinced that the type of matter the authors consider to estimate such a scale is not applicable (at least) at all cosmological times, thus leading to incorrect or inconclusive general conclusions, and predictions.

\subsection{Standard model Lagrangian and matter Lagrangian}

As previously mentioned, one of the biggest conceptual problems in~\cite{Muk} is the idea that in order to scrutinize cosmological models and the implications of the non-minimal coupling, commonly regarded and treated as a classical interaction between geometry and matter in cosmology, it is possible to assign the Standard Model Lagrangian $L_\text{SM}$ to the matter Lagrangian $L_\text{m}$ at all scales. However, it is crucial to bear in mind that modified gravity theories with a non-minimal geometry-matter coupling are classical field theories and not quantum field theories, despite the possible manifestation of certain quantum features within them.

Usually, in schematic terms, the action for gravitational theories, including the standard GR, is given by
\begin{equation}
\label{classical_gravity}
    S=\int{\left(\text{Classical Gravity}+\text{Classical Matter}\right)\sqrt{-g}d^4x},
\end{equation}
However, in the case of $f(R, \rm{Matter})$ type theories, the `Classical Gravity" term (may) include a non-minimal coupling. Hence, more generally, the action of the modified gravity theories of this class can be formulated as
\begin{equation}
\label{classical_gravity1}
    S=\int{f\left(\text{Classical Gravity},\text{Classical Matter}\right)\sqrt{-g}d^4x},
\end{equation}
where $f$ is an arbitrary analytical function of its variables.

 Moreover, at this moment, one may ask the following question: since any term in Eqs.~\eqref{classical_gravity} and (\ref{classical_gravity1}) is classical, can we \textit{realistically} include a quantum field theory formulated in Minkowski space-time (in this case, the SM) in it, and expect \textit{accurate cosmological predictions}?

  On the one hand, on the most fundamental level (in terms of length scales) and at almost every cosmological era (in terms of energy scales), our best current theory for describing matter is the Standard Model of particle physics. On the other hand, we know that it is not easy to ``marry" gravity (and, therefore, cosmology) with the Standard Model, as they are incompatible at most of the cosmological time, not only due to a matter of length scales but also due to technical problems (the Standard Model precisely works when gravity is switched ``off"). Moreover, the matter we deal with in cosmology (although at the quantum fundamental level described by the Standard Model) is a classical fluid in which the fundamental aspects of the Standard Model do not play any relevant role due to the emergent, thermodynamic character of the fluid. Thus, the conflict between the nature of the Standard Model and the nature of Eqs. \eqref{classical_gravity} and \eqref{classical_gravity1} means that the issue on the energy scales is quite subtle and, as such, must be treated with care.

In truth, it is impossible for one even to say that it is achievable to put a classical and a quantum field theory on equal feet, otherwise, the problem of quantum gravity would have been probably solved by now. What one can argue is that, in a first-order approximation, to grasp possible impacts, such an inclusion is valid. Nevertheless, it can not tell the complete story. 

For example, although some of the analysis carried out in Ref.~\cite{Muk} is mathematically correct, such as the one in the subsection \textbf{Non-linear coupling function}, in which a specific model of $f(R,T)$ gravity was constrained by using $W$ bosons, it raises some important conceptual questions regarding its physical validity.
In particular, when the authors conclude that ``the coupling $1/2 \gamma_2 T^2$ is irrelevant for any process whose energy density is below $m^4_W$, hence the theory loses its interest for e.g cosmology today'' \cite{Muk} (with $\gamma_2$ being a constant), it can be misleading because they are doing a first-order approximation in a specific energy scale in which it is possible to do so, at least in principle: right after the electroweak phase transition, that is, somewhere between the vacuum expectation value $v \approx 246$ GeV and 100 GeV. Other than that, it is purely speculative to do such a treatment. In fact, according to the Standard Model itself, all gauge bosons were massless before the electroweak phase transition. Hence, it is not possible to constrain $f(R, T)$ as done by the authors of \cite{Muk} before this symmetry breaking due to the absence of a mass term, which demonstrates the lack of universality of the approach. Moreover, after a sufficient amount of time, the Standard Model Lagrangian starts to be irrelevant to analyzing cosmological models. Let us address this issue in more detail.

\subsubsection{During/Right After Electroweak Phase Transition}

The authors in Ref. \cite{Muk} used the Standard Model of particle physics to rule out, for late-time cosmological applications, some $f(R, T)$ gravity models. In the following, we will be presenting and discussing their line of reasoning.

When the electroweak phase transition occurs, the  maximal internal symmetry of the Standard Model Lagrangian is spontaneously broken as
\begin{equation}
    SU(2)_W \times U(1)_Y \longrightarrow U(1)_Q,
\end{equation}
 where $W$, $Y$ and $Q$ refer to weak isospin, weak hypercharge, and electric charge, respectively. It is well-known that it is this symmetry breaking that allows the $Z^0$ and $W$ bosons to acquire mass via the Higgs mechanism, which leads to the terms used by the authors
\begin{eqnarray}
\label{lm_post}
L_\text{m} \supset-\frac{1}{2}\left(\partial_\mu W_\nu^{+}-\partial_\nu W_\mu^{+}\right)\left(\partial^\mu W_{-}^\nu-\partial^\nu W_{-}^\mu\right)
+\frac{m_W^2}{2} W_\mu^{+} W_{-}^\mu,
\end{eqnarray}

Then, the authors of \cite{Muk} compute the corresponding trace of Eq. \eqref{lm_post}, which they denote by $T_W$. Since the only term that survives in $T_W$ is the mass term, the expression for $1/2 \gamma_2 T^2_W$ is \cite{Muk}
\begin{equation}
    \frac{1}{2} \gamma_2 T^2_W = 2 \gamma_2 m^4_W \left(W^{+}_{\mu} W_{-}^{\mu}\right)^2 .
    \label{T_W^2}
\end{equation}
As correctly pointed out in \cite{Muk}, the above expression is quartic in $W$, and since the only quartic terms in $W$ present in the Standard Model are
\begin{equation}
    \label{3.14 Muk}
    L_{\rm{SM}} \supset -\frac{g^2}{2}\left[\left(W_\mu^{+} W_{-}^\mu\right)^2-W_\mu^{+} W_{+}^\mu W_\nu^{-} W_{-}^\nu\right],
\end{equation}
with $g$ being the electroweak coupling constant, this means that it is not possible to absorb  Eq. \eqref{T_W^2} into the Standard Model. Accordingly, we have contributions from new $f(R,T)$ gravity physics. Due to partial wave unitarity, and taking into account some experimental constraints, the authors obtain
\begin{equation}
\left|\gamma_2 m_W^4\right| \leq 0.05.
\end{equation}
According to the authors of \cite{Muk}, with $m_W \simeq  80$ GeV, the term $1/2 \gamma_2 T^2$ would be negligible or irrelevant up to $m^4_W \simeq 4.1 \times 10^7 $ GeV. However, as we are going to see, this may not be the case.

Nonetheless, this analysis is done by assuming that $L_\text{m} \supset L_\text{SM}$ and that $L_\text{SM}$ is always relevant for all cosmological (length and energy) scales. Indeed, based on a physically questionable assumption, the authors exclude any contribution of $T^2$ and higher terms for late-time cosmology. We agree that such an exclusion might ``make sense'' during/right after electroweak scales, but it can not be a \textit{universal} prediction. For instance, at the dark ages, such a term ($T^2$) can be relevant as in this cosmological era one can not make the same assignment and do the first-order approximation anymore, leaving, in an effective manner, the $L_\text{SM}$ out of $L_\text{m}$.

\subsubsection{Before Electroweak Phase Transition}
In order to refute the argument shown above, let us apply the exact same reasoning to moments before the electroweak phase transition.

It is known that before the electroweak phase transition, the Standard Model (electroweak) Lagrangian has its maximal internal symmetry, that is $SU(2)_W \times U(1)_Y$, and all Standard Model elementary particles are in thermal equilibrium. Furthermore, the massless electroweak gauge bosons are $W_{1}$, $W_{2}$, $W_{3}$ and $B$. As such, the matter Lagrangian (which contains the Standard Model Lagrangian) contains the following free terms for the electroweak gauge boson fields $W_{1}$ and $W_{2}$
\begin{equation}
\label{lm_before}
L_\text{m} \supset -\frac{1}{4} \left(W_1^{\mu \nu} W_{\mu \nu}^1 +  W_2^{\mu \nu} W_{\mu \nu}^2 \right),
\end{equation}
where $W_1^{\mu \nu}$ and $W_2^{\mu \nu}$ are the strength field tensor associated with $W_{1}$ and $W_{2}$ bosons, respectively. Moreover, these can be expressed as
\begin{equation}
    W_1^{\mu \nu}=\partial^\mu W_1^\nu-\partial^\nu W_1^\mu+g W_2^\mu W_3^\nu-g W_3^\mu W_2^\nu,
\end{equation}
and
\begin{equation}
W_2^{\mu \nu}=\partial^\mu W_2^\nu-\partial^\nu W_2^\mu+g W_3^\mu W_1^\nu-g W_1^\mu W_3^\nu.
\end{equation}
Thus, due to the absence of a mass term in Eq. \eqref{lm_before}, the corresponding trace is obviously 0. Therefore, it is not possible to constrain $f(R, T)$ gravity by using the method that was previously shown, which implies that the initial prediction is not, at least, entirely valid. As such, we might have a possible contribution from $T^2$ and higher order terms (coming from beyond the Standard Model exotic imperfect fluids or some conformal anomaly) between $m^4_W$ and electroweak scales that can not be probed by particle physics experiments.

\subsubsection{``Most of the time"}
Although we recognize that \textit{on a quantum level} $L_\text{m}$ should include the SM Lagrangian, its assignment to the (classical) matter Lagrangian is not valid for \textit{all} scales. In particular, for non-minimal geometry-matter coupling cosmology, the matter that it is considered is the cosmological matter in the form of a perfect fluid (with $L_\text{m} = p$ or $L_\text{m} = - \rho$); it applies to the universe as a whole. As a result, the energy density and the pressure  $\rho$ and $p$ of the perfect fluid should not be thought of as they are constructed from another (scalar, vector, ...) field. Furthermore, the prediction that a $T^2$ is only relevant at very high energy scales may not occur for these Lagrangians, the ones adopted for late-time cosmology as they depict better the matter content of the universe in these cosmological times due to their emergent, thermodynamic nature.

To summarize this Section, we verify that the prediction ``the coupling $1/2 \gamma_2 T^2$ is irrelevant for any process whose energy density is below $m^4_W$, hence the theory loses its interest for e.g cosmology today'' is highly dependent on the matter Lagrangian (and, therefore, on the cosmological time), and, because of that, one can not state that the prediction is universal in the sense that it can not be valid for all cosmological times. Additionally, we state two examples of conceptual shortcomings regarding the analysis done in Ref.~\cite{Muk}
\begin{itemize}
\item If $W$ bosons are being described by classical fields, then the analysis is not physically correct, as $W$ bosons do not have a classical field counterpart (the range of the weak interaction is $r_W=\hbar / m_W c \simeq 2.5 \times 10^{-16} \mathrm{~cm}$, with $m_W \simeq 80 \text{GeV}$ \cite{Altarelli:2000ma});
\item If $W$ bosons are treated as quantum fields then the potential problem of the compatibility between the nature of both theories (SM and modified gravity) arises.
\end{itemize}

Indeed, the issue on the energy scales in the context of cosmologies with a non-minimal geometry-matter coupling is both relevant and interesting, but we think that the analysis done in \cite{Muk} is not entirely correct due to the arguments above. Nonetheless, we recognize that near curvature singularities, where these theories might be a first-order approximation for a quantum theory of the gravitational interaction (in which Eqs. \eqref{act} and \eqref{123} might describe such dynamics), the inclusion of the Standard Model Lagrangian is mandatory and therefore the \textit{modus operandi} adopted in \cite{Muk} may be completely valid if it is somehow adapted to the curved-spacetime (and, thus, not using the Minkowski metric). Furthermore, one can also see this exchange of arguments as the pursuit of a more correct criterion to evaluate a possible energy scale for the contributions of non-minimal geometry-matter coupling in cosmology.

\section{$f(R,\text{Matter})$ theories, and cosmological observations}\label{sect4}

It is indeed absolutely correct to asses the viability of a given gravitational theory through its compatibility with the observations \cite{Muk}. In this Section, we will investigate the cosmological implications of a rather involved example of a $f(R,\text{Matter})$ type theory, containing a complicated coupling between matter and geometry. We will explicitly show that the cosmological observational data can be well described with these type of theories.

\subsection{Action and field equations}

Let us consider an action functional of the form
\begin{align}
	S=\int d^4x\sqrt{-g}\left[\kappa^2 (R+f(R,L_m))+L_m\right],
\end{align}
where $L_m$ is the baryonic matter Lagrangian. In this section, we will consider a special case for the function $f$ as
\begin{align}
	f(R,L_m)=\alpha R|L_m|^\beta,
\end{align}
where $\alpha$ and $\beta$ are some constants.

The equation of motion of the model can be obtained by varying the corresponding action with respect to the metric, yielding
\begin{align}\label{metric}
	G_{\mu\nu}+f_RR_{\mu\nu}-\frac12fg_{\mu\nu}+(g_{\mu\nu}\Box-\nabla_\mu\nabla_\nu)f_R=\frac{1}{2\kappa^2}(1+\kappa^2f_L)T_{\mu\nu}+\frac12f_LL_mg_{\mu\nu},
\end{align}
where subscripts $R$ and $L$ represent derivative with respect to $R$ and $L_m$. 

In general, in theories with non-minimal matter-geometry couplings, the energy-momentum tensor is not conserved. Taking the covariant divergence of the metric field equation \eqref{metric}, on obtains the (non)-conservation equation as
\begin{align}\label{cons}
	\nabla^\mu T_{\mu\nu}=\frac{\kappa^2}{1+\kappa^2f_L}(L_mg_{\mu\nu}-T_{\mu\nu})\nabla^\mu f_L.
\end{align}
One can see from the above relation that for $f_L=const.$ the energy-momentum tensor becomes conserved. This is the case where there is no matter-geometry couplings in the theory.

For this special case of matter-geometry coupling theory, a very interesting situation happens. Suppose that the universe can be described by a Friedmann-Lemaitre-Robertson-Walker metric, with line element
\begin{align}
	ds^2 = -dt^2+a(t)^2(dx^2+dy^2+dz^2),
\end{align}
where $a(t)$ is the scale factor. We also assume that the matter content of the universe can be described by a perfect fluid with matter Lagrangian $L_m$ and the energy-momentum tensor
\begin{align}
	T_{\mu\nu}=(\rho+p)u_\mu u_\nu+pg_{\mu\nu},
\end{align}
where $\rho$ is the thermodynamic energy density of the cosmic fluid and $p$ is its pressure. In this case, the conservation equation \eqref{cons} reduced to
\begin{align}
	\dot\rho+3H(\rho+p)=\frac{\kappa^2}{1+\kappa^2 f_L}(\rho+L_m)\dot{f}_L.
\end{align}
One can see that irrelevant to the functional relation of $f$, the energy-momentum tensor is conserved for $L_m=-\rho$. As a result, in this section, we will use this choice to have a conservation matter-geometry coupling theory. It should be noted that this only happens in the background level; the non-conservative nature of the theory will show up at the perturbation level.

\subsection{The generalized Friedmann equations}
Using the above assumptions, we obtain the generalized Friedmann equation of this model as
\begin{align}
3H^2=\frac{1}{2\kappa^2}\rho-3\alpha H^2\rho^\beta+9\alpha\beta(1+\omega)H^2\rho^\beta,
\end{align}
where $\omega=p/\rho$ is the equation of state parameter.

In the following, we will consider {\it the late-time cosmological evolution of the Universe}, and try to fit the model with a small set of observational Hubble data. Consequently,  we will assume that the Universe is filled with dust, with equation of state of the form $\omega=0$.

\paragraph{Dimensionless variables and redshift representation.} After defining the following set of dimensionless variables
\begin{align}
	\tau =H_0t, \quad H=H_0 h,\quad \bar{\rho}=\frac{\rho}{6\kappa^2H_0^2},\quad
\bar{\alpha}=\frac{6^\beta\kappa^{2\beta}H_0^{2\beta}}{2}\alpha,
\end{align}
and transforming to the redshift representation, with the redshift $z$ defined as
\begin{align}
	1+z=\frac1a,
\end{align}
the rescaled Hubble parameter takes the form
\begin{align}
	h(z)=\sqrt{\frac{3\Omega_{m0}(1+z)^3}{3-\bar\alpha(3\beta-1)\Omega_{m0}^\beta(1+z)^{3\beta}}},
\end{align}
where $\Omega_{m0}$ is the matter density abundance at the present time $(z=0)$. Remembering that the energy-momentum tensor is conserved, the behavior of the matter density abundance can be obtained as
\begin{align}
	\Omega_m(z)=\frac{\Omega_{m0}(1+z)^3}{h^2},
\end{align} 
which is the same as the behavior we have seen in standard GR.
\subsection{Statistical analysis, and comparison with observations}

In order to find the best fit value of the parameters $H_0$, $\Omega_{m0}$, $\bar\alpha$ and  $\beta$,  we use the Likelihood analysis using the observational data on the Hubble parameter in the redshift range $z\in(0.07,2.36)$  tabulated in \cite{hubble1}.

In the case of independent data points, the likelihood function can be defined as
\begin{align}\label{68}
	\mathcal{L}=\mathcal{L}_0e^{-\chi^2/2},
\end{align}
where $\mathcal{L}_0$ is the normalization constant, and the quantity $\chi^2$ is defined as
\begin{align}\label{69}
	\chi^2=\sum_i\left(\frac{O_i-T_i}{\sigma_i}\right)^2.
\end{align}
Here $i$ counts the data points, $O_i$ are the observational value, $T_i$ are the theoretical values, and $\sigma_i$ are the errors associated with the $i$th data obtained from observations.

By maximizing the likelihood function, the best fit values of the parameters $H_0$, $\Omega_{m0}$, $\bar\alpha$ and $\beta$ at $1\sigma$ confidence level, can be obtained as
\begin{figure*}
	\includegraphics[scale=0.5]{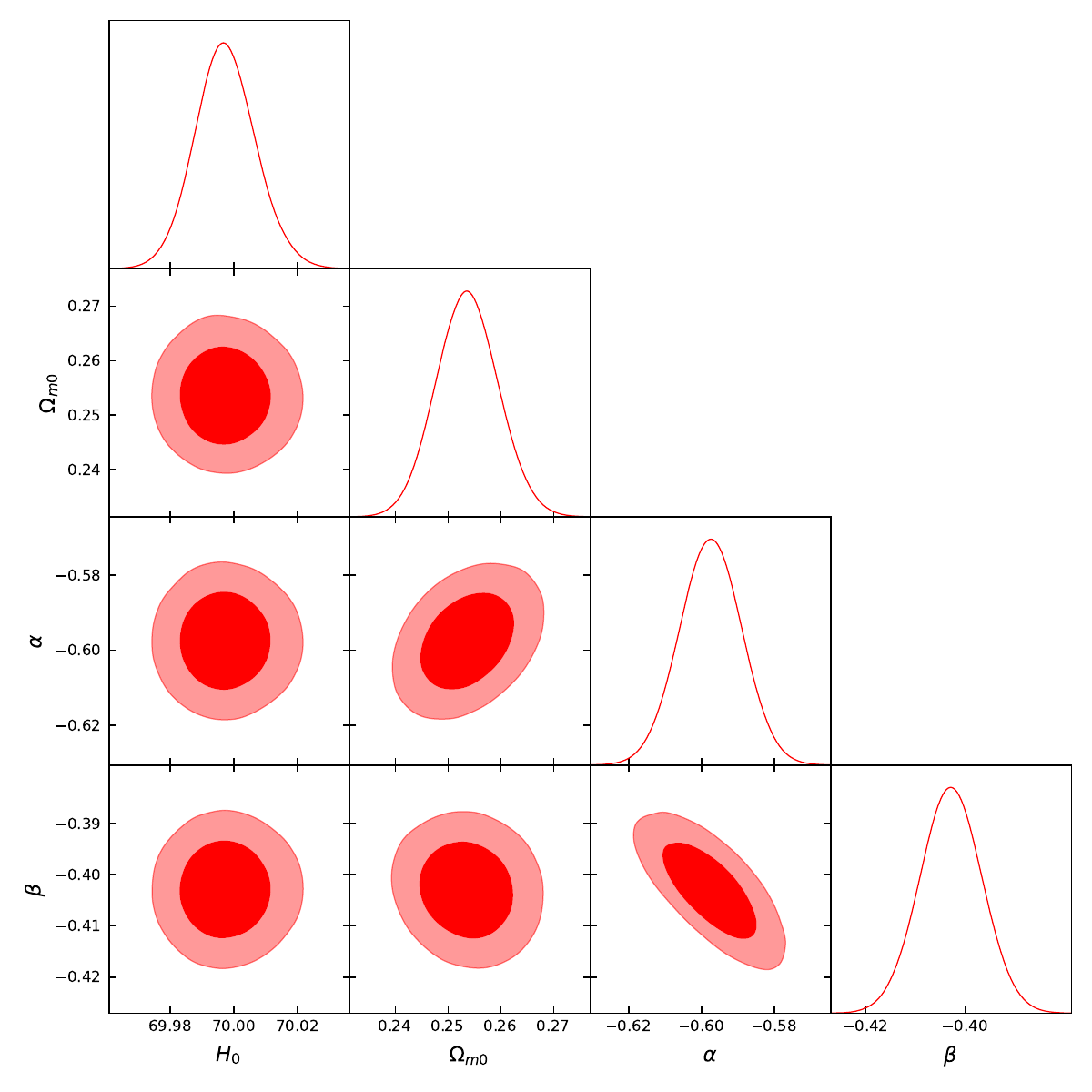}
	\caption{The corner plot for the values of the parameters $H_0$, $\Omega_{m0}$, $\bar\alpha$ and $\beta$ with their $1\sigma$ and $2\sigma$ confidence levels. \label{cornerplot}}
\end{figure*}

\begin{align}\label{bestfit}
	H_0&=69.997^{+0.0096}_{-0.0092},\nonumber\\
	\Omega_{m0}&=0.254^{+0.0059}_{-0.0058},\nonumber\\
	\bar\alpha&=-0.597^{+0.0084}_{-0.0085},\nonumber\\
	\beta&=-0.403^{+0.0062}_{-0.0062}.
\end{align}
The corner plot for the values of the parameters  $H_0$, $\Omega_{m0}$, $\bar\alpha$ and $\beta$ with their $1\sigma$ and $2\sigma$ confidence levels is shown in Fig. \ref{cornerplot}.

\begin{figure*}[htbp]
	\includegraphics[scale=0.57]{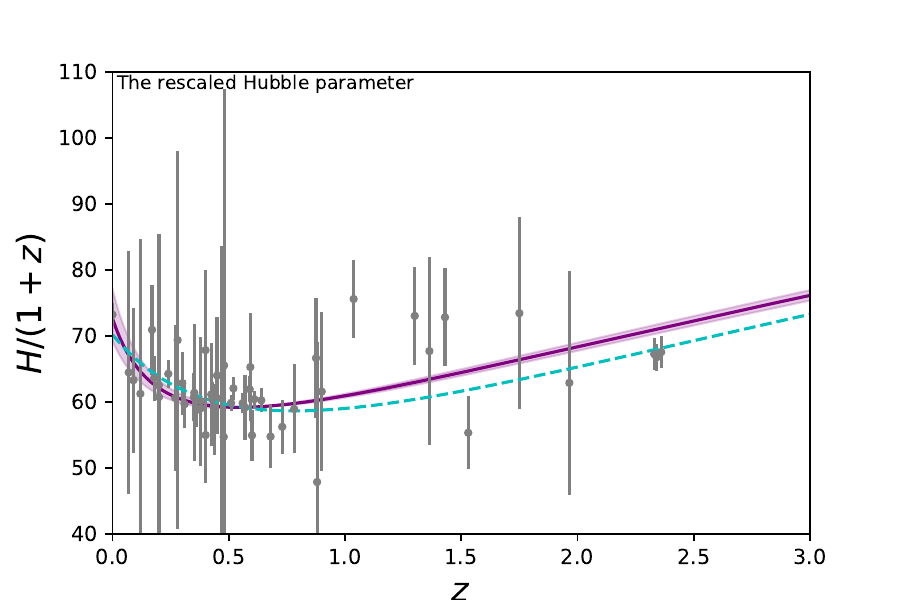}\includegraphics[scale=0.57]{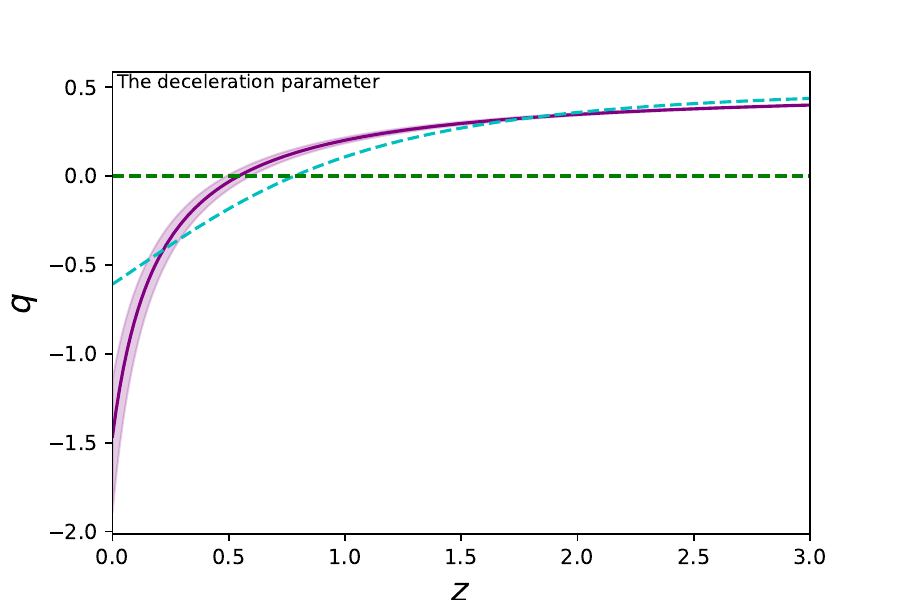}
	\caption{\label{fighubq} The behavior of the rescaled Hubble parameter $H(z)/(1+z)$ (left panel) and of the deceleration parameter $q(z)$ (right panel) as a function of the redshift $z$ for the best fit values of the parameters of the $f(R,{\rm Matter})$ cosmological model, as given by Eqs.~(\ref{bestfit}). The shaded area denotes the $1\sigma$ error. The dashed line represents the predictions of the $\Lambda$CDM model.}
\end{figure*}

By using the optimal values of the model parameters we can study now the cosmological evolution as predicted by the model. The behaviors of the Hubble parameter and of the deceleration parameter are represented, as functions of $z$, in Fig.~\ref{fighubq}. As one can see from this Figure, the present cosmological model provides a good description of the observational data, where its predictions are very close to the predictions of the $\Lambda$CDM model. However, small differences between the predictions of the two models will possibly be observationally significant. Differences between models also exist in the case of the deceleration parameter (see the right panel of Fig.~\ref{fighubq}). This property of the model may give some clear observational evidence that may help in discriminating between modified gravity models with geometry-matter coupling, and the standard cosmological paradigm.

\section{Conclusions}\label{sect5}

The main goal of the present work was twofold. Firstly, we have presented a critical assessment of the analysis performed in \cite{Muk} on the physical foundations, and applications, of the $f(R,{\rm Matter})$ type theories. Secondly, we have presented the theoretical basics and justifications of these types of approaches to the gravitational interaction. The main reason for proposing such an extension of standard GR is not difficult to find, or to explain, and it is fully related to the interesting, and challenging situation in present-day cosmology. Around 25 years ago, an observational fact was discovered, namely, that the Universe is undergoing an accelerating expansion, a fact that drastically changed our views on gravity and cosmology.  This result forced theorists (and even observational cosmologists) to find answers to the new fundamental questions on the nature of gravity emerging from the increased number of new findings on the structure and dynamics of the Universe.

The most basic theoretical question, still needing an answer, is if GR, based on the Einstein-Hilbert action plus a matter term, $S=\int{\left(R/2\kappa ^2+L_\text{m}\right)\sqrt{-g}d^4x}$, is the correct relativistic theory of gravitation. Presently, GR (assumed to describe gravity in the standard $\Lambda$CDM cosmological model) is facing many challenges related to the difficulty of explaining recent observations, from which we would like to mention, for example, the Hubble and the related tensions \cite{Val}, and the incompatibility with quantum mechanics, the other well-established physical theory. All these observational and theoretical aspects seem to indicate that GR breaks down at large astrophysical/cosmological scales, going beyond the scale of the Solar System, and (possibly/very probably) at microscopical scales.

Hence, the need for new gravitational physics is well-motivated by the present status of the field. Since gravitational field theories are essentially based on a variational principle, the most straightforward way to extend GR is to consider a more general action describing gravity. Among others, these new theories may offer the possibility of a more realistic representation of the gravitational interaction near curvature singularities (and perhaps avoiding them), and they may lead to at least some first-order approximations for a (possible) quantum theory of gravity. From a theoretical point of view one of the most interesting questions is related to the possibility, or existence, of a maximal extension of the Einstein-Hilbert Lagrangian density. This Lagrangian density has a simple additive structure, so as a first step towards its generalization one could consider different algebraic structures, including multiplicative ones. These extensions would automatically lead to $f(R,{\rm Matter})$ type theories. Moreover, some hints in the direction of the possible physical relevance of this kind of approach may be provided by some results in quantum gravity, or quantum field theory in curved spacetimes, where geometry-matter coupling, or particle creation, does naturally occur. But of course the validity of these approaches to gravity must not be taken as granted, and in this context any critical discussion on the foundations, applications, viability or relevance of $f(R,{\rm Matter})$ type theories  is very much welcomed. That is why the paper \cite{Muk} is a useful contribution in this field, opening a critical debate on the very foundations of gravitational theories, and especially on those involving a geometry-matter coupling. However, even if the goal, and intentions of  \cite{Muk} are positive in a general scientific context, the overall approach, results and criticism of the $f(R,{\rm Matter})$ gravity raises a number of questions, and they cannot be considered as an (even partial) answer to the intricate problem of the viability of these theories. In the present paper we have considered, and discussed in detail, all of the criticism presented in \cite{Muk}. The most problematic aspect of the analysis in \cite{Muk} is the definition, or interpretation, of the concept of matter, which in $f(R,{\rm Matter})$ theory is considered as the standard matter, consisting of protons, neutrons, electrons etc. To explain the cosmological observations, $f(R,{\rm Matter})$ gravity does not need scalar, Higgs, or other vector fields, since its main goal is to build up a cosmological/astrophysical theory that is directly testable by using the easily detectable forms of matter. In short, the arguments presented in \cite{Muk} are not relevant for astrophysics/cosmology, but from them the authors still infer that these theories are not relevant for these fields.

Moreover, it is the role of the baryonic matter only that is reevaluated within the theory. Hence, a similar discussion as the one performed in \cite{Muk}, but focused on the presence and effects of the baryonic matter, described by a thermodynamic density and  pressure only, would lead indeed to a deeper understanding of the viability and relevance of this class of gravitational theories.

An additional relevant consideration introduced in \cite{Muk} is related to the existence of an energy scale beyond which $f(R,\text{Matter})$ type theories are no longer valid. Indeed, we started to discuss this issue, commonly present in the effective field theory approach to high-energy particle physics, by mentioning that the relevant scales in classical modified gravity theories are length scales and not energy scales. In fact, the relation between the two is not trivial, and more investigation should be carried out. But even if one assumes that it is possible to assess these theories in this manner, we have demonstrated that the reasoning applied in \cite{Muk} to establish an energy scale to a particular $f(R,T)$ model is either not universal or suffers from severe conceptual physical problems.

The basic criterion of the acceptance/rejection of a physical theory is its capability of explaining the observational data. From this point of view, $f(R,{\rm Matter})$ theories do not perform particularly badly, and they have been investigated extensively in a cosmological/astrophysical context. We have presented a particular cosmological model, based on an extended Einstein-Hilbert action, containing an extra  geometry-matter coupling term of the form $\left|L_m\right|^\beta R$. Such a term in the Lagrangian gives a good description of the observational data for the Hubble function, and reproduces the predictions of the $\Lambda$CDM model without resorting to a cosmological constant, or to scalar and vector fields.

Hence, given the arguments presented in this paper, we strongly believe that $f(R,\rm{Matter})$ type theories can shed light on the interesting challenges present-day gravitational physics and cosmology are facing.

\end{document}